\documentclass[aps,reprint,groupedaddress,nofootinbib]{revtex4-1}
\usepackage{dcolumn}
\usepackage{graphicx}
\usepackage{color}

\newcounter{firstbib}

\begin{document}

\title{Molecules cooled below the Doppler limit}

\author{S. Truppe, H. J. Williams, M. Hambach, L. Caldwell, N. J. Fitch, E. A. Hinds, B. E. Sauer and M. R. Tarbutt}
\email[]{m.tarbutt@imperial.ac.uk}
\affiliation{Centre for Cold Matter, Blackett Laboratory, Imperial College London, Prince Consort Road, London SW7 2AZ, UK}

\maketitle

The ability to cool atoms below the Doppler limit -- the minimum temperature reachable by Doppler cooling -- has been essential to most experiments with quantum degenerate gases, optical lattices and atomic fountains, among many other applications. A broad set of new applications await ultracold molecules~\cite{Carr2009}, and the extension of laser cooling to molecules has begun~\cite{Shuman2010, Hummon2013, Zhelyazkova2014, Hemmerling2016, Kozyryev2016}. A molecular magneto-optical trap has been demonstrated~\cite{Barry2014, McCarron2015, Norrgard2016}, where molecules approached the Doppler limit. However, the sub-Doppler temperatures required for most applications have not yet been reached. Here we cool molecules to 50~$\mu$K, well below the Doppler limit, using a three-dimensional optical molasses. These ultracold molecules could be loaded into optical tweezers to trap arbitrary arrays~\cite{Barredo2016} for quantum simulation~\cite{Micheli2006}, launched into a molecular fountain~\cite{Tarbutt2013b, Cheng2016} for testing fundamental physics~\cite{Hudson2011, Baron2014, Truppe2013}, and used to study ultracold collisions and ultracold chemistry~\cite{Krems2008}.

Sub-Doppler cooling usually occurs in a light field with non-uniform polarization, where optical pumping between sub-levels sets up a friction force at low velocity much stronger than the Doppler force~\cite{Dalibard1989, Ungar1989}. Details of the polarization-gradient cooling mechanism depend on the ground- and excited-state angular momenta $F''$ and $F'$ for the relevant transitions. In our case, where $F'' \ge F'$ (see Methods), there are dark ground states that cannot couple to the local polarization of the light, and bright states that can. A bright-state molecule loses kinetic energy on moving into blue-detuned light, and pumps to a dark state only when the light has sufficient intensity. The now dark molecule moves through changing polarization, where it switches non-adiabatically back to a bright state, preferentially at low light intensity where the energies of dark and bright states are similar. Thus, the molecule continually loses kinetic energy. This method~\cite{Weidemuller1994, Boiron1995}, often called  ``grey molasses'', has been used to cool atoms. Magnetically-induced laser cooling~\cite{Ungar1989} involves a similar mechanism but uses a suitable magnetic field instead of the polarization gradient so that the Larmor precession transfers molecules between the dark and bright sub-levels. This mechanism has been used to cool molecules in one dimension~\cite{Shuman2010, Kozyryev2016} but sub-Doppler temperatures were not reached.

Almost all ultracold atom experiments begin with a magneto-optical trap (MOT), which is likely to become a workhorse for cooling molecules too. Until now, only SrF molecules have been trapped this way. For these molecules, two types of MOT have been developed, a dc MOT~\cite{Barry2014, McCarron2015} and a radio-frequency (rf) MOT where optical pumping into dark states is avoided by rapidly reversing the magnetic field and the handedness of the MOT laser~\cite{Norrgard2016}. Our experiment begins with a dc MOT of CaF, prepared by methods similar to those of ~\cite{Barry2014, McCarron2015} (see Methods). A pulse of CaF molecules produced at time $t=0$ is emitted from a cryogenic buffer gas source and decelerated to about 15~m/s by frequency-chirped counter-propagating laser light. The main slowing laser is denoted ${\cal L}^{{\rm s}}_{00}$. The slow molecules are captured in the MOT between $t=16$ and 40~ms. The main MOT laser, denoted ${\cal L}_{00}$, drives a transition of linewidth $\Gamma = 2\pi\times 8.3$~MHz.
The magnetic quadrupole field $B$ has an axial gradient of 2.9~mT/cm, and the background magnetic field is adjusted to maximize the number of trapped molecules using shim coils along each axis.

\begin{figure*}[tb]
	\includegraphics[width=1.7\columnwidth]{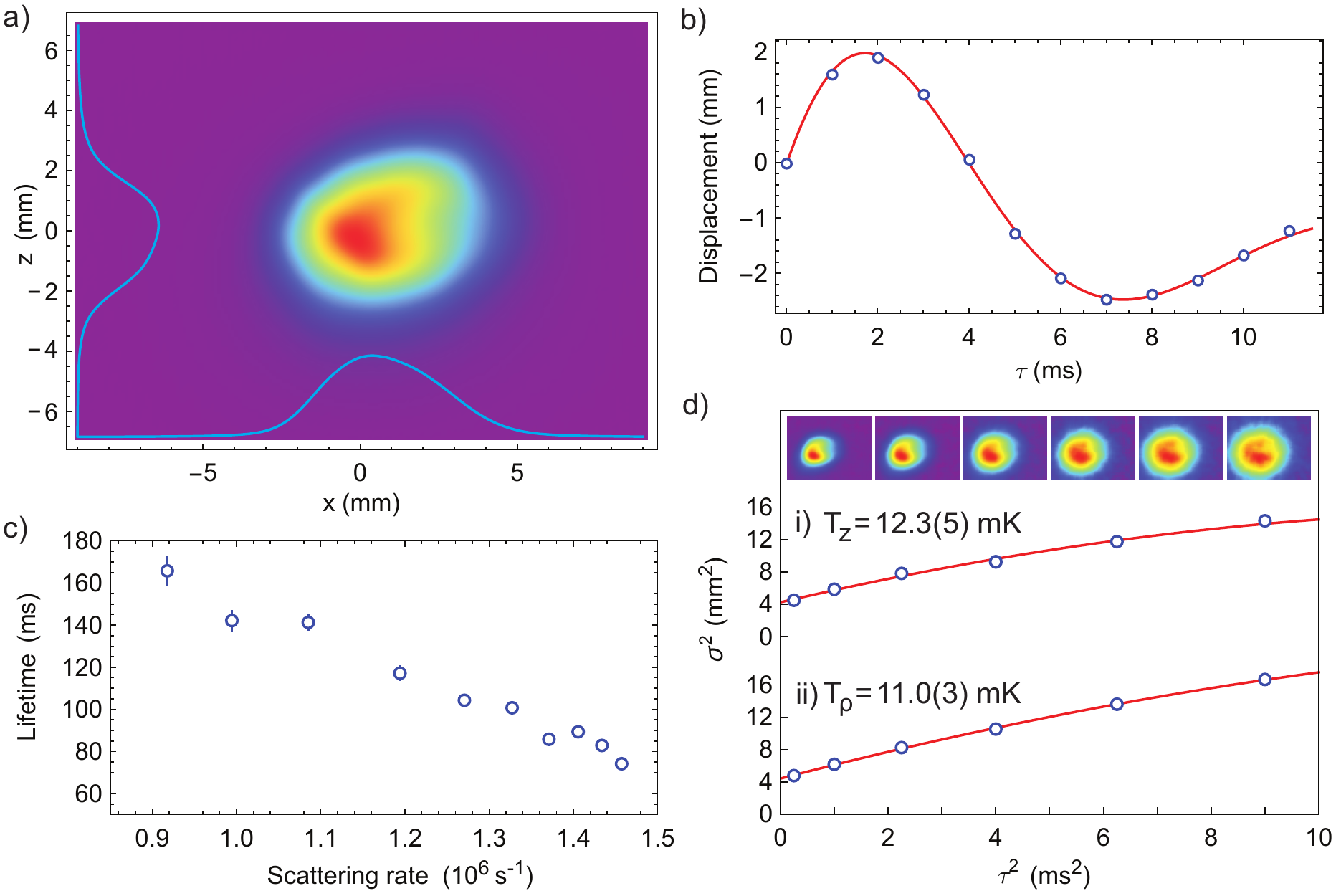}
	\caption{{\bf Characterisation of the MOT.} a) Fluorescence image of the MOT, averaged over 100 shots with 50~ms exposure starting at $t=40$~ms. Also shown are the 1D axial and radial density profiles obtained by integrating the image along each axis. b) Radial displacement of the cloud versus time after pushing it. Each point is obtained by summing 50 images each of 1~ms exposure. Solid line: fit to the motion of a damped harmonic oscillator. c) MOT lifetime versus scattering rate. The scattering rate is controlled by changing the intensity of ${\cal L}_{00}$, and is inferred from the measured decay time of the MOT fluorescence after switching off ${\cal L}_{21}$ (see Methods). d) Temperature measurement using the free expansion method. The mean squared width of the cloud is plotted against the square of the free expansion time, for  i) the axial and ii) the radial directions. Each point is obtained by summing 50 images each of 1~ms exposure. The top row shows typical images. Solid lines: quadratic fit to the data (see Methods). Where error bars are not visible in b), c), d), they are smaller than the point size.\label{MOT}}
\end{figure*}

Figure 1(a) shows the molecules in the MOT, imaged on a CCD camera by collecting their fluorescence from $t=40$~ms to $t=90$~ms. We estimate that there are $1.8(5) \times 10^{4}$ molecules in this MOT (see Methods), with a peak density of $n=2.3(6)\times 10^{5}$~cm$^{-3}$. The MOT disappears when $B$ is reversed or turned off. As the frequency of ${\cal L}_{00}$ is increased, there comes a critical frequency where MOT loading becomes unstable. We introduce $\Delta$, the detuning of ${\cal L}_{00}$, and define $\Delta=0$ to be at this critical frequency. We observe MOTs when $0 > \Delta > -1.8\Gamma$, and we load the most molecules when $\Delta = -0.75\Gamma$, the value used for all data in this paper. Integrating a CCD image along principal axes, we obtain 1D axial and radial density profiles, as shown in Fig.~1(a). Gaussian fits yield the axial and radial centres and rms widths. These are used to obtain oscillation frequencies, damping constants and temperatures. To measure the trap's radial oscillation frequency and damping constant, we push the molecules radially using a 500~$\mu$s pulse of ${\cal L}^{{\rm s}}_{00}$ light at $t=50$~ms, then image them for 1~ms after a fixed delay, $\tau$. Figure 1(b) shows $\rho(\tau)$, the mean radial displacement of the molecules as a function of this delay. Describing $\rho(\tau)$ by the damped harmonic oscillator equation, $\rho'' + \beta \rho' + \omega^{2} \rho = 0$, we determine an oscillation frequency of $\omega = 2\pi \times 94.4(2)$~Hz and a damping constant of $\beta = 390(4)$~s$^{-1}$. Similar values were found in SrF MOTs~\cite{McCarron2015, Norrgard2016}. To determine the MOT lifetime, we fit the decay of its fluorescence to a single exponential. Figure 1(c) shows this lifetime as a function of the scattering rate, varied by changing the intensity of ${\cal L}_{00}$ and measured as described in Methods. We see that the lifetime, typically 100~ms, decreases with higher scattering rate, suggesting loss by optical pumping to a state that is not addressed by the lasers. We do not see the precipitous drop in lifetime observed at low scattering rate in the dc MOT of SrF~\cite{Norrgard2016}.

To measure the temperature we turn off $B$ and ${\cal L}_{00}$, then turn ${\cal L}_{00}$ back on  after a delay time $\tau$ to image the cloud using a 1~ms exposure. From the image, we determine the mean squared widths $\sigma^{2}$ in the axial and radial directions. These are plotted against $\tau^{2}$ in Fig.~1(d), together with fits to the model described in Methods. These fits give an axial temperature of $T_{z}=12.3(5)$~mK and a radial temperature of $T_{\rho}=11.0(3)$~mK, and this $\sim 10\%$ variation is typical of all our data. We choose to  present temperatures as $T=T_{\rho}^{2/3}T_{z}^{1/3}$, giving  $T=11.4(3)$~mK in this case. The corresponding phase space density is $\rho=h^3 n/(2\pi m k_B T)^{3/2}=2.2(6)\times10^{-15}$. 

\begin{figure}[h]
	\includegraphics[width=\columnwidth]{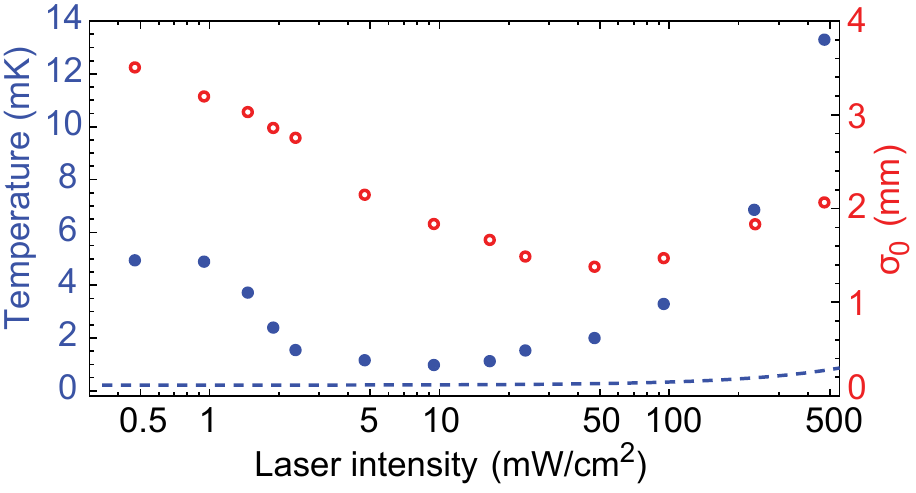}
	\caption{{\bf Cooling by ramping down the laser intensity}. Temperature (blue filled circles) and geometric mean rms width of the MOT (red open circles) vs total intensity of ${\cal L}_{00}$ at the end of the ramp. Dashed line is the Doppler temperature given by Eq.~(\ref{Eq:DopplerLimit}). Error bars are smaller than the points. \label{TvsI}}
\end{figure}

 \begin{figure*}[tb]
 	\includegraphics[width=1.8\columnwidth]{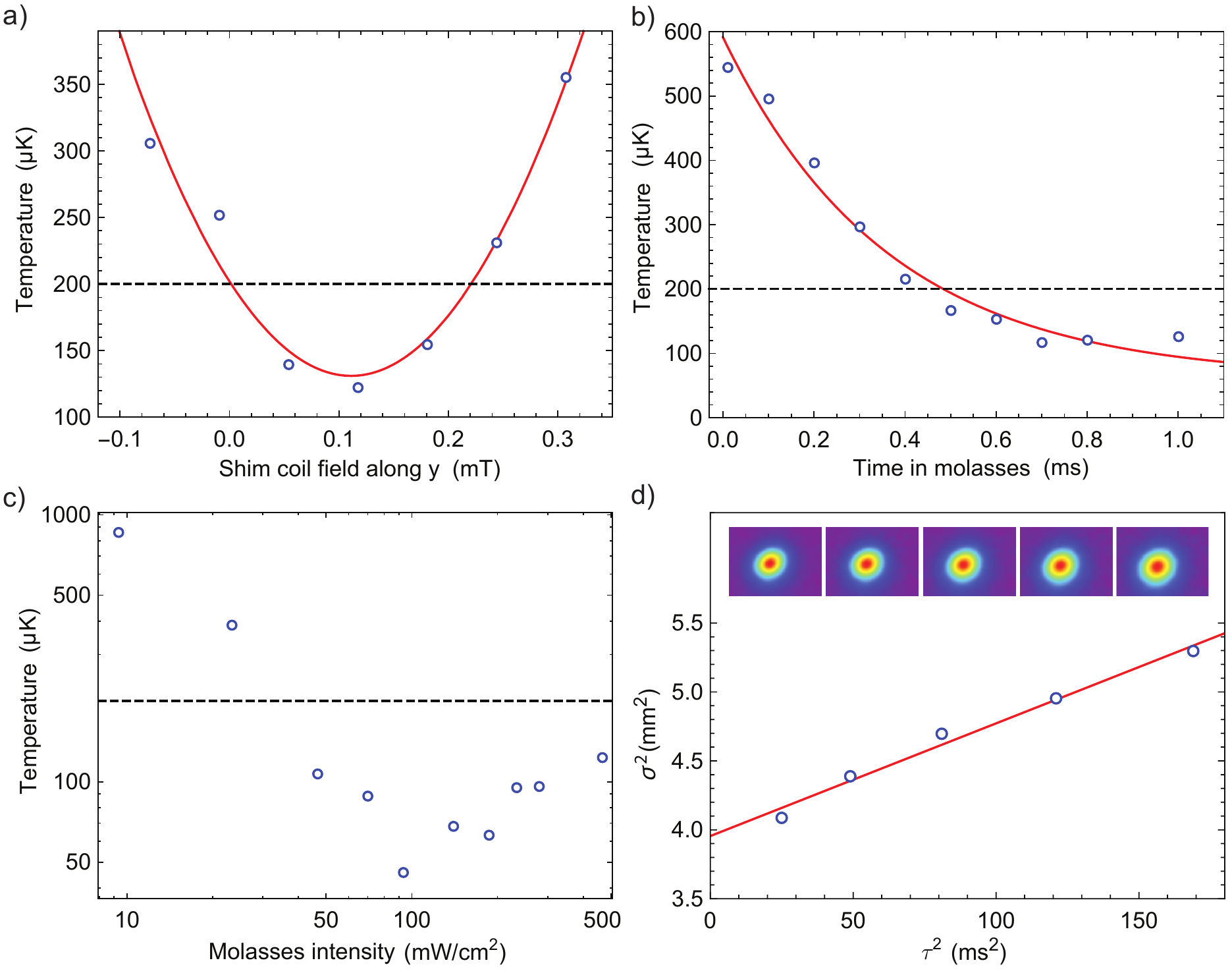}
 	\caption{{\bf Sub-Doppler cooling}. (a) Temperature versus magnetic field produced at the cloud by one of three shim coils. The intensity of ${\cal L}_{00}$ is 460~mW/cm$^{2}$ and the molecules are held in the molasses for 5~ms. Solid line is a quadratic fit and has a curvature of 5740(30)~$\mu$K/mT$^{2}$. (b) Temperature versus time in the molasses. The intensity of ${\cal L}_{00}$ is 460~mW/cm$^{2}$. Solid line is a fit to an exponential decay, giving a $1/e$ time constant of 361(2)~$\mu$s. Dashed line shows the minimum Doppler temperature. (c) Temperature versus ${\cal L}_{00}$ intensity during the molasses phase. The molasses is on for 5~ms. (d) Free expansion temperature measurement after a period of 5~ms in a molasses. The intensity of ${\cal L}_{00}$ is 93~mW/cm$^{2}$. The inset shows the images from which the cloud size at each time point is obtained. The solid line is a straight line fit. From 5 repeated measurements we obtain a temperature of 52(2)~$\mu$K. \label{Molasses}}
 \end{figure*}

We expect the Doppler temperature to be 
\begin{equation}\label{Eq:DopplerLimit}
T_{\rm D} = -\frac{\hbar\Gamma^{2}}{8 k_{\rm B} \Delta}(1+ s_{\rm eff} + 4\Delta^{2}/\Gamma^{2}),
\end{equation}
where $s_{\rm eff} = I_{00}/I_{\rm s,eff}$ is an effective saturation parameter, $I_{00}$ is the total intensity at the MOT from ${\cal L}_{00}$, and $I_{\rm s,eff} \approx 50$~mW/cm$^{2}$ is an effective saturation intensity (see Methods). $T_{\rm D}$ has its minimum value $T_{\rm D,min} = \hbar \Gamma \sqrt{1+s_{\rm eff}} / (2 k_{\rm B}) = 200\mu{\rm K}\sqrt{1+s_{\rm eff}}$ at detuning $\Delta = -\Gamma/2 \sqrt{1+s_{\rm eff}}$. With our parameters ($s_{\rm eff}=9.3$, $\Delta=-0.75\Gamma$) Eq.~(\ref{Eq:DopplerLimit}) gives $T_{\rm D} = 830$~$\mu$K, 14 times lower than the measured value discussed above. The elevated temperature at high intensity is similar to those observed in SrF MOTs~\cite{Barry2014, McCarron2015, Norrgard2016}. This may be due to a balance between Doppler forces and polarization-gradient forces, which together drive the molecules towards a non-zero equilibrium speed whose value increases with intensity~\cite{Devlin2016}.

For a hot rf MOT of SrF, lowering the laser intensity reduced the temperature to 400~$\mu$K without loss of molecules~\cite{Norrgard2016} and then to 250~$\mu$K~\cite{Steinecker2016}, but at substantial cost to the number and density.
This method was not useful for cooling the dc MOT of SrF because of its short lifetime at low intensity. By contrast, the lifetime of our dc CaF MOT increases at lower intensity, so this method is open to us. We decrease the power in ${\cal L}_{00}$ between $t=50$ and 70~ms, hold it for 5~ms, then measure the MOT temperature as described above. Figure~\ref{TvsI} shows both the temperature and the size of the MOT versus final intensity, together with $T_D$ given by Eq.~(\ref{Eq:DopplerLimit}). At 9.2~mW/cm$^{2}$ we find a minimum of 960~$\mu$K, about 4 times the value of $T_{\rm D}$. Ramping to lower intensities increases the temperature again. Optimization of $\Delta$ and the shim coils lowers the minimum temperature to about 500~$\mu$K, but we do not pursue that further here. The MOT size first decreases as the intensity decreases, but then grows once the intensity is below 50~mW/cm$^{2}$. 
After ramping down to 9.2~mW/cm$^{2}$ the cloud has $n=1.1(3) \times 10^{5}$~cm$^{-3}$ and $\rho=4.3(1.1) \times 10^{-14}$.

Next, we transfer the molecules into a three-dimensional blue-detuned optical molasses. We ramp down ${\cal L}_{00}$ (as above) to 4.6~mW/cm$^{2}$, and hold that intensity until $t=76$~ms. We switch the shim coil currents to new values at $t=72$~ms, and switch off the MOT coils at $t=75$~ms. At $t=76$~ms the detuning is switched to $\Delta = +2.5\Gamma$ to make the molasses, and ${\cal L}_{00}$ is switched to a (variable) higher intensity, both within 10~$\mu$s. After a brief hold time, typically 5~ms, we measure the temperature by our usual method with $\Delta$ restored to $-0.75\Gamma$ for the imaging step. Figure~\ref{Molasses} shows how this temperature depends on the key parameters. The temperature is sensitive to all three components of the magnetic field.  Figure \ref{Molasses}(a) shows the quadratic variation of temperature versus one field component after optimising the other two. Figure \ref{Molasses}(b) shows the temperature in the molasses evolving towards a base value of $\sim 100$~$\mu$K with a rapid $1/e$ time constant of 361(2)~$\mu$s. In Fig.~\ref{Molasses}(c) we show the temperature versus the intensity of ${\cal L}_{00}$ during the molasses phase. The temperature has a minimum of 46~$\mu$K near 100~mW/cm$^{2}$, increases rapidly at lower intensities and more gradually at higher intensities. The temperature dependencies shown in Fig.~\ref{Molasses}(a-c) are all similar to those observed in atomic grey molasses~\cite{Fernandes2012, Boiron1995}. Figure \ref{Molasses}(d) shows the thermal expansion of a cloud after cooling for 5~ms in a 100~mW/cm$^{2}$ molasses. The average of 5 such temperature measurements gives $T=52(2)~\mu$K. To within our 5\% uncertainty, no molecules are lost between the initial MOT and this ultracold cloud, excepting loss due to the MOT lifetime. The cloud now has $n=1.1(3) \times 10^{5}$~cm$^{-3}$ and $\rho=3.4(9) \times 10^{-12}$, 1500 times higher than in the initial MOT.

Single molecules from this ultracold gas could be loaded into low-lying motional states of microscopic optical tweezer traps and formed into regular arrays~\cite{Barredo2016} for quantum simulation~\cite{Micheli2006}. They could be loaded into chip-based electric traps and coupled to transmission line resonators, forming the elements of a quantum processor~\cite{Andre2006}. By mixing the molecules with atoms, it will be possible to explore collisions, chemistry~\cite{Krems2008} and sympathetic cooling~\cite{Lim2015} in the ultracold regime. Our cooled molecules could be used to search for a time variation of the electron-to-proton mass ratio~\cite{Kajita2009}, while application of the methods to other amenable molecules will advance measurements of electric dipole moments~\cite{Tarbutt2013b, Hunter2012} and nuclear anapole moments~\cite{Cahn2014}.  Major increases in density are likely to come from more efficient slowing methods~\cite{Fitch2016} along with transverse cooling~\cite{Shuman2010} prior to slowing. The resulting dense, ultracold sample is an ideal starting point for sympathetic or evaporative cooling to quantum degeneracy.

{\bf Acknowledgements.} We thank Jack Devlin for his assistance and insight. We are grateful to Jon Dyne, Giovanni Marinaro and Valerijus Gerulis for technical assistance. The research has received funding from EPSRC under grants EP/I012044 and EP/M027716, and from the European Research Council under the European Union's Seventh Framework Programme (FP7/2007-2013) / ERC grant agreement 320789.

{\bf Author contributions.} All authors made substantial contributions to this work.

{\bf Competing financial interests.} The authors declare no competing financial interests.


\section{Methods}

\begin{figure}[tb]
	\includegraphics[width=\columnwidth]{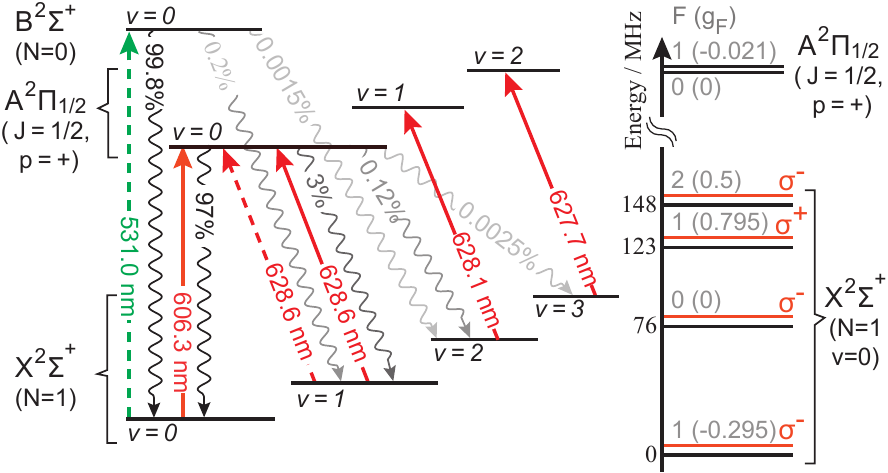}
	\caption{{\bf Relevant energy levels of CaF.} The main slowing transition is ${\rm B}^{2}\Sigma^{+} (v=0,N=0) \leftarrow {\rm X}^{2}\Sigma^{+} (v=0,N=1)$. The main MOT transition is ${\rm A}^{2}\Pi_{1/2} (v=0,J=1/2,p=+) \leftarrow {\rm X}^{2}\Sigma^{+} (v=0,N=1)$. All repumping is done by driving ${\rm A}^{2}\Pi_{1/2} (v-1,J=1/2,p=+) \leftarrow {\rm X}^{2}\Sigma^{+} (v,N=1)$ transitions. Dashed lines are transitions driven by the slowing lasers. Solid lines are transitions driven by the MOT lasers. Wavy lines indicate decay channels with branching ratios given. The right panel shows the hyperfine structure of the main MOT transition. The ground state has four resolved hyperfine components with $F=1,0,1,2$, while the excited state has two unresolved components with $F=0,1$. The magnetic $g_{F}$-factors are also shown. The relative polarization handedness of the laser frequency components driving each hyperfine component are indicated as $\sigma^{\pm}$. The labels $v, N, J, F, p$ are the quantum numbers of vibration, rotational angular momentum, total electronic angular momentum, total angular momentum and parity.\label{levels}}
\end{figure}

{\bf Laser cooling scheme.} Figure \ref{levels} shows the energy levels in CaF relevant to the experiment, and the branching ratios between them. The two excited states, ${\rm A}^{2}\Pi_{1/2}$ and ${\rm B}^{2}\Sigma^{+}$, have decay rates of $\Gamma=2\pi\times 8.3$~MHz~\cite{Wall2008} and $2\pi\times 6.3$~MHz~\cite{Dagdigian1974} respectively. The main slowing laser (${\cal L}^{{\rm s}}_{00}$) drives the ${\rm B}^{2}\Sigma^{+}(v'=0) \leftarrow {\rm X}^{2}\Sigma^{+}(v''=0)$ transition at 531.0~nm. Population that leaks to $v''=1$ during the slowing is returned to the cooling cycle by a repump slowing laser (${\cal L}^{{\rm s}}_{10}$) that drives the ${\rm A}^{2}\Pi_{1/2}(v'=0) \leftarrow {\rm X}^{2}\Sigma^{+}(v''=1)$ transition at 628.6~nm. The MOT uses four lasers, denoted ${\cal L}_{ij}$, to drive the ${\rm A}^{2}\Pi_{1/2}(v'=j) \leftarrow {\rm X}^{2}\Sigma^{+}(v''=i)$ transitions. These are ${\cal L}_{00}$ at 606.3~nm, ${\cal L}_{10}$ at 628.6~nm, ${\cal L}_{21}$ at 628.1~nm and ${\cal L}_{32}$ at 627.7~nm. All lasers drive the P(1) component so that rotational branching is forbidden by the electric dipole selection rules~\cite{Stuhl2008}. Each of the levels of the X state shown in Figure~\ref{levels} is split into four components due to the spin-rotation and hyperfine interactions. The splittings for the $v''=0$ state are shown in the figure, while those for the other states are similar. Radio-frequency (rf) sidebands are added to each laser (see discussion of Fig.~\ref{sidebands} below) to ensure that all these components are addressed. The frequency component of ${\cal L}_{00}$ that addresses the upper $F=1$ state has the opposite circular polarization to the other three. Thus, when the overall detuning of ${\cal L}_{00}$ is negative, as in Fig.~\ref{levels}, the $F=2$ component is driven simultaneously by two frequencies with opposite circular polarization, one red- and the other blue-detuned. This configuration produces the dual-frequency MOT described in~\cite{Tarbutt2015}. The simulations presented there suggest that most of the confining force in the MOT is due to this dual-frequency effect.

\begin{figure}[tb]
	\includegraphics[width=\columnwidth]{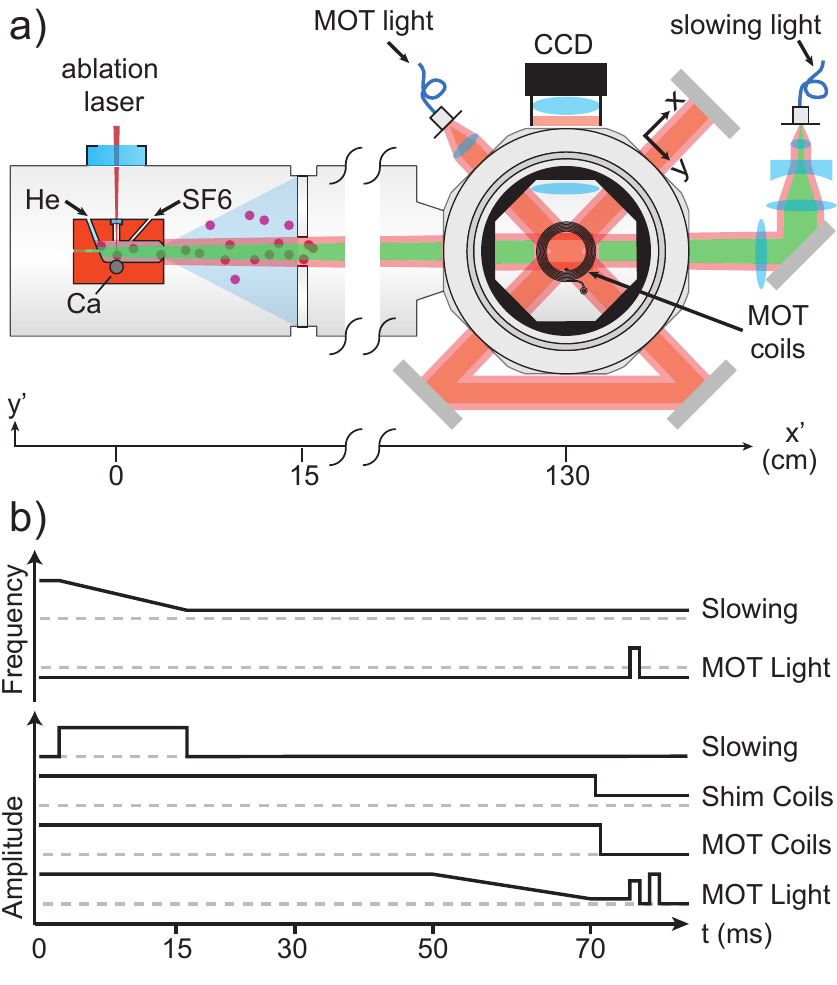}
	\caption{{\bf Schematic of the experiment}. a) Apparatus. b) Timing diagram. Upper graph shows how the frequencies of ${\cal L}^{{\rm s}}_{00}$ and ${\cal L}_{00}$ change in time. Lower graph shows how the powers of ${\cal L}^{{\rm s}}_{00}$ and ${\cal L}_{00}$ and the currents in the shim coils and MOT coils change in time.\label{setup}}
\end{figure}

{\bf Setup and procedures.} Figure \ref{setup}(a) illustrates the experiment. A short pulse of CaF molecules is produced at $t=0$ by laser ablation of a Ca target in the presence of SF$_6$. These molecules are entrained in a continuous 0.5~sccm flow of helium gas cooled to 4~K, producing a pulsed beam with a typical mean forward velocity of 150~m/s. The beam exits the source through a 3.5~mm diameter aperture at $x'=0$, passes into the slowing chamber through an 8~mm diameter aperture at $x'=15$~cm, and then through a 20~mm diameter, 200~mm long differential pumping tube whose entrance is at $x'=90$~cm, reaching the MOT at $x'=130$~cm. The pressure in the slowing chamber is $6 \times 10^{-8}$~mbar, and in the MOT chamber is $2 \times 10^{-9}$~mbar. The experiment runs at a repetition rate of 2~Hz.

The beam is slowed using the methods described in \cite{Truppe2016}. The slowing light is combined into a single beam, containing 100~mW of ${\cal L}^{{\rm s}}_{00}$ and 100~mW of ${\cal L}^{{\rm s}}_{10}$. This beam has a $1/e^{2}$ radius of 9~mm at the MOT, converging to 1.5~mm at the source. The changes in frequency and intensity of ${\cal L}^{{\rm s}}_{00}$ are illustrated in the timing diagram in Fig.~\ref{setup}(b). The initial frequency of  ${\cal L}^{{\rm s}}_{00}$ is set to a detuning of -375\,MHz so that molecules moving at 200~m/s are Doppler-shifted into resonance. The light is switched on at $t=2.5$~ms and frequency chirped at a rate of 23~MHz/ms between 3.4~ms and 15~ms. The frequency of  ${\cal L}^{{\rm s}}_{10}$ is not chirped, which differs from the procedure used previously~\cite{Truppe2016}. Instead, it is frequency broadened as described below, and its centre frequency detuned by 200\,MHz. Both ${\cal L}^{{\rm s}}_{00}$ and ${\cal L}^{{\rm s}}_{10}$ are turned off at $t=15$~ms. A 0.5~mT magnetic field, directed along $y'$, is applied throughout the slowing region and is constantly on.

The MOT light is combined into a single beam containing 80~mW of ${\cal L}_{00}$,  100~mW of ${\cal L}_{10}$,  10~mW of ${\cal L}_{21}$ and 0.5~mW of ${\cal L}_{32}$. This beam is expanded to a $1/e^{2}$ radius of 8.1~mm, and then passed through the centre of the MOT chamber six times, first along $y$, then $x$, then $z$, then $-z$, then $-x$, then $-y$. In this paper, intensity refers always to the six-beam intensity of ${\cal L}_{00}$. The light is circularly polarized each time it enters the chamber, and returned to linear polarization each time it exits, following \cite{Barry2014}. For any given frequency component of the light, the handedness is the same for each pass in the horizontal plane, but opposite in the vertical direction. All MOT lasers have zero detuning, apart from ${\cal L}_{00}$ which has variable detuning $\Delta$. The MOT field gradient, which is 2.9~mT/cm in the axial direction, is produced by a pair of anti-Helmholtz coils inside the vacuum chamber. Three bias coils, with axes along $x'$, $y'$ and $z$, are used to tune the magnetic field in the MOT region to trap the most molecules. The MOT fluorescence at 606~nm is collected by a lens inside the vacuum chamber and imaged onto a CCD camera with a magnification of 0.5. An interference filter blocks background light at other wavelengths. 

The complete procedure for cooling to the lowest temperatures is illustrated in Fig.~\ref{setup}(b). The intensity of ${\cal L}_{00}$ is ramped down by a factor 100 to 4.6~mW/cm$^{2}$ between $t=50$ and 70~ms to lower the MOT temperature (see Fig.~\ref{TvsI}), while keeping the detuning at $-0.75\Gamma$. At $t=72$~ms, the shim coil currents are switched from those that load the most molecules in the MOT to those that give the lowest temperature in the molasses, thereby optimising both molecule number and temperature. The MOT coils are turned off at $t=75$~ms. At $t=76$~ms we jump ${\cal L}_{00}$ to a detuning of $+2.5\Gamma$, and to a (variable)  higher intensity, to form the molasses (see Fig.~\ref{Molasses}(c)). After allowing the molasses to act for a variable time (see Fig.~\ref{Molasses}(b)) ${\cal L}_{00}$ is turned off so that the cloud can expand for a variable time (see Fig.~\ref{Molasses}(d)) before it is imaged for 1~ms at full intensity with $\Delta=-0.75\Gamma$.  

Figure~\ref{sidebands} shows the frequency spectrum of each laser. For ${\cal L}_{00}$, a 73.5~MHz electro-optic modulator (EOM) generates the sidebands that drive the $F=2$, $F=0$ and lower $F=1$ states, while a 48~MHz acousto-optic modulator (AOM) generates the light of opposite polarization to address the upper $F=1$ state~\cite{Zhelyazkova2014}. The rf sidebands for ${\cal L}_{10}$, ${\cal L}_{21}$, ${\cal L}_{32}$ and ${\cal L}^{{\rm s}}_{00}$ are generated using 24~MHz EOMs. We spectrally broaden ${\cal L}^{{\rm s}}_{10}$ to approximately 500~\,MHz using three consecutive EOMs, one driven at 72~MHz, one at 24~MHz and one at 8~MHz. For each laser, we find the frequency that maximizes the laser-induced fluorescence (LIF) from the molecular beam when that laser is used as an orthogonal probe. These frequencies define zero detuning for each laser, with the exception of ${\cal L}_{00}$. For ${\cal L}_{00}$, we find that there is a critical frequency where an observable MOT is only formed in half of all shots.  We define this critical frequency to be zero detuning, $\Delta = 0$. At $\Delta = -2$~MHz the MOT is stable, and at $\Delta = 2$~MHz there is never a MOT. When ${\cal L}_{00}$ is used as an orthogonal probe, the LIF is maximized at $\Delta = 2\pi \times 2(4)$~MHz.

\begin{figure}[tb]
	\includegraphics[width=\columnwidth]{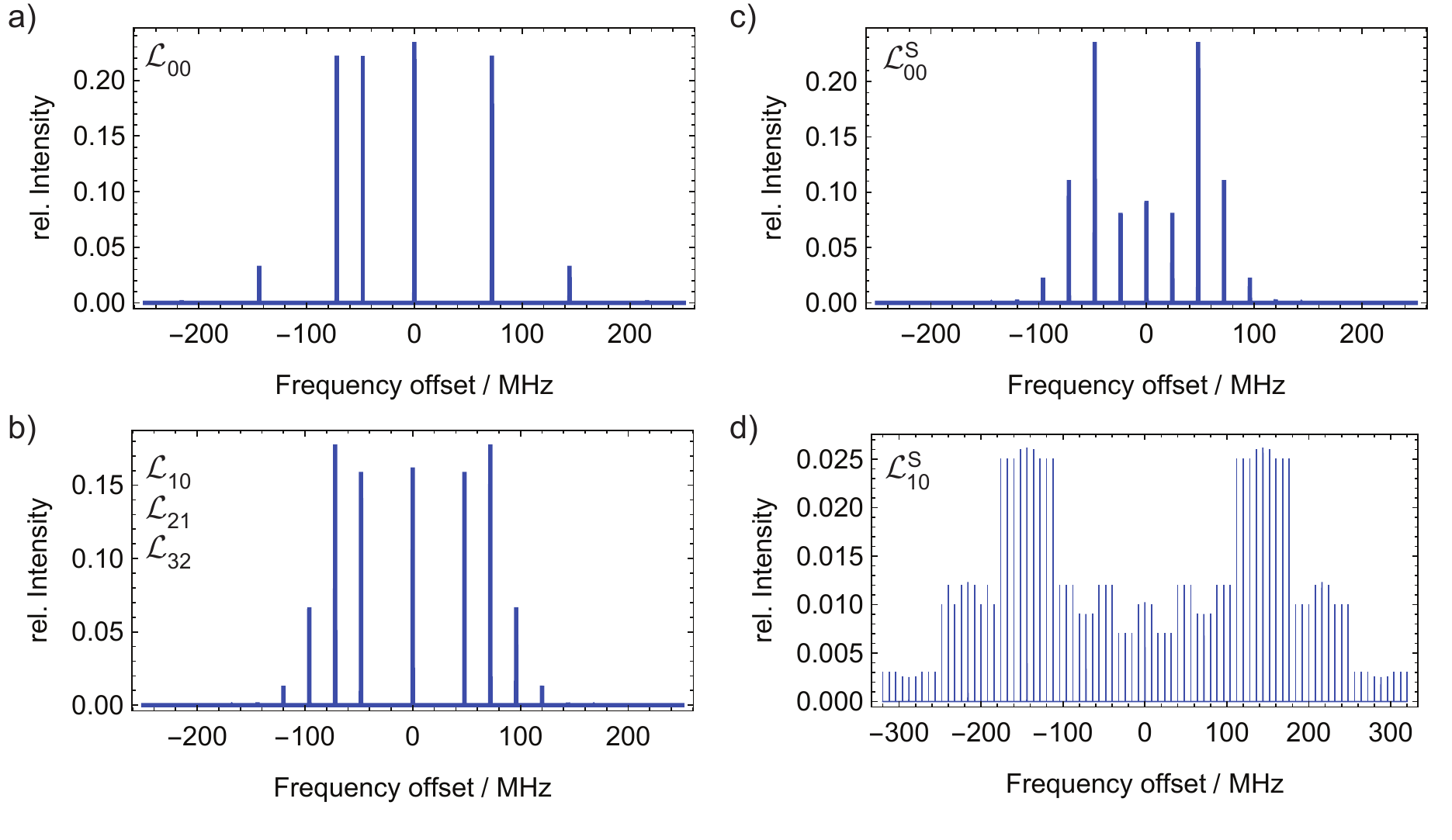}
	\caption{{\bf Frequency spectrum of each laser.} a) ${\cal L}_{00}$. b)  ${\cal L}_{10}$,  ${\cal L}_{21}$,  and ${\cal L}_{32}$. c) ${\cal L}^{{\rm s}}_{00}$. d) ${\cal L}^{{\rm s}}_{10}$. \label{sidebands}}
\end{figure}

{\bf Scattering rate and saturation intensity.} Despite the complexity of the multi-level molecule, it is useful to use a simple rate model~\cite{Tarbutt2013b} to predict some of the properties of the MOT, as done previously~\cite{Norrgard2016}. In this model, $n_g$ ground states are coupled to $n_e$ excited states, and the steady-state scattering rate is found to be
\begin{equation}
R_{\rm{sc}} = \Gamma \frac{n_e}{(n_g + n_e) + 2\sum_{j=1}^{n_{g}}(1+4\Delta_{j}^{2}/\Gamma^{2})I_{\rm{s},j}/I_{j}}.
\label{eq:Rsc1}
\end{equation}
Here, $I_j$ is the intensity of the light driving transition $j$, $\Delta_j$ is its detuning, and $I_{\rm{s},j}=\pi h c \Gamma/(3 \lambda_{j}^{3})$ is the two-level saturation intensity for a transition of wavelength $\lambda_{j}$. In applying this model we need to include the $n_g=24$ Zeeman sub-levels of the $v=0$ and $v=1$ ground states, all of which are coupled to the same $n_{e}=4$ levels of the excited state. The $v=2$ and $v=3$ ground states can be neglected since they are repumped through other excited states with sufficient intensity that their populations are always small. Because ${\cal L}_{00}$ is detuned whereas ${\cal L}_{10}$ is not, and because the ${\cal L}_{10}$ intensity is always higher than the ${\cal L}_{00}$ intensity, the transitions driven by ${\cal L}_{10}$ make only a small contribution to the sum in Eq.~(\ref{eq:Rsc1}) and we neglect them. This is a reasonable approximation at full ${\cal L}_{00}$ power, and a very good approximation once the power of ${\cal L}_{00}$ is ramped down. The 12 transitions driven by ${\cal L}_{00}$ have common values for $\Delta$ and $I_{\rm{s}}$, and the total intensity, $I_{00}$, is divided roughly equally between them so that we can write $I_{j}=I_{00}/(n_{g}/2)$. With these simplifications, we can rewrite Eq.~(\ref{eq:Rsc1}) in the form 
\begin{equation}
\label{eq:Rsc2}
R_{\rm{sc}} = \frac{\Gamma_{\rm{eff}}}{2} \frac{s_{\rm{eff}}}{1 + s_{\rm{eff}} + 4\Delta^{2}/\Gamma^{2}},
\end{equation}
where
\begin{equation}
\label{eq:GammaEff}
\Gamma_{\rm{eff}} = \frac{2 n_{e}}{n_{g} + n_{e}} \Gamma = \frac{2}{7}\Gamma,
\end{equation}
and
\begin{equation}
\label{eq:sEff}
s_{\rm_{eff}} = \frac{2(n_{g} + n_{e})}{n_{g}^{2}} \frac{I_{00}}{I_{\rm{s}}}.
\end{equation}
Writing $s_{\rm eff} = I_{00}/I_{\rm s,eff}$, we find an effective saturation intensity of
\begin{equation}
\label{eq:Is}
I_{\rm s,eff} = \frac{n_{g}^{2}}{2(n_{g} + n_{e})} I_{\rm{s}} = \frac{72}{7}  I_{\rm{s}} = 50~{\rm mW/cm}^{2}.
\end{equation}

We have measured the scattering rate at various ${\cal L}_{00}$ intensities, using the method described in the next paragraph. These measurements show that the scattering rate does indeed follow the form of Eq.~(\ref{eq:Rsc2}), but suggest that $\Gamma_{\rm{eff}}$ is roughly a factor of 3 smaller than Eq.(\ref{eq:GammaEff}) while $I_{\rm s,eff}$ is roughly a factor of 2 smaller than Eq.~(\ref{eq:Is}). However, the determination of $I_{\rm s,eff}$ is sensitive to the value of the detuning which is imperfectly defined for the multi-level molecule. Fortunately, none of our conclusions depend strongly on knowing the values of either $\Gamma_{\rm{eff}}$ or $I_{\rm s,eff}$.  

{\bf Molecule number.} To estimate the number of molecules in the MOT, we need to know the photon scattering rate per molecule. We measure this by switching off ${\cal L}_{21}$ and recording the decay of the fluorescence as molecules are optically pumped into $v''=2$. The decay is exponential with a time constant of 570(10)~$\mu$s at full ${\cal L}_{00}$ intensity. Combining this with the branching ratio of 0.12\% to $v''=2$~\cite{Pelegrini2005} gives a scattering rate of $1.5\times 10^{6}$~s$^{-1}$. This is about 3.5 times below the value predicted by Eq.~(\ref{eq:Rsc2}). The detection efficiency is 1.6(2)\% and is determined by numerical ray tracing together with the measured transmission of the optics and the specified quantum efficiency of the camera. For the MOT shown in Fig.~\ref{MOT}(a), the detected photon count rate at the camera is $4.3 \times 10^{8}$~s$^{-1}$. From these values, we estimate that there are $1.8(2) \times 10^{4}$ molecules in this MOT. From one day to the next, using nominally identical parameters, and after optimization of the source, the molecule number varies by about 25\%. We assign this uncertainty to all molecule number estimates in this paper.

{\bf Temperature.} In the standard theory of Doppler cooling, the equilibrium temperature is reached when the Doppler cooling rate equals the heating rate due to the randomness of photon scattering. Because both rates are proportional to the scattering rate, the multi-level system is expected to have the same Doppler-limited temperature as a simple two-level system. This is the temperature given by Eq.~(\ref{Eq:DopplerLimit}).

We measure the temperature using the standard ballistic expansion method. For a thermal velocity distribution and an initial Gaussian density distribution of rms width $\sigma_{0}$, the density distribution after a free expansion time $\tau$ is a Gaussian with a mean squared width given by $\sigma^{2} = \sigma_{0}^{2} + k_{\rm B} T \tau^{2}/m$, where $m$ is the mass of the molecule. Thus, a plot of $\sigma^{2}$ against $\tau^{2}$ should be a straight line whose gradient gives the temperature. 

There are several potential sources of systematic error in this measurement which we address here. First we consider whether the finite exposure time of 1~ms introduces any systematic error. While the image is being taken using the MOT light, the magnetic quadrupole field is off. Thus, there is no trapping force, but there is a velocity-dependent force which, according to  \cite{Devlin2016}, may either accelerate or decelerate the molecules depending on whether their velocity is above or below some critical value. To quantify the effect, we have made temperature measurements using various exposure times. For a 12~mK cloud we estimate that the 1~ms exposure time results in an overestimate of the temperature by about 0.3(5)~mK. For a 50~$\mu$K cloud, the overestimate is about 0(3)~$\mu$K. These corrections are insignificant.

With a MOT that is centred on the light beams, the intensity of the imaging light is higher in the middle of the cloud than it is in the wings, making the cloud look artificially small. This skews the temperature towards lower values because the effect is stronger for clouds that have expanded. The error is mitigated by using laser beams that are considerably larger than the cloud and that strongly saturate the rate of fluorescence. Using a three-dimensional model of the MOT beams and Eqs.~(\ref{eq:Rsc2}) and (\ref{eq:Is}) for the dependence of the scattering rate on intensity, we have simulated the imaging to determine the functional form of $\sigma^{2}(\tau^{2})$ expected in our experiment. The model suggests that a simple $ \tau^{4}$ correction -- $\sigma^{2} = \sigma_{0}^{2} + k_{\rm B} T \tau^{2}/M + a_{2} \tau^{4}$ -- will fit well to all our ballistic expansion data, will recover the correct temperature, and will give a significantly non-zero $a_{2}$ for our $T \sim 10$~mK data, but a negligible one for all our data where $T \le 1$~mK. We have investigated this in detail using $\sigma^{2}$ versus $\tau^{2}$ data with a higher density of data points (12 points, instead of our usual 6). For a hot MOT ($T \sim 12$~mK), we see a non-linear expansion and find that a fit to the above ``quadratic model'' gives a temperature that is typically about 10\% higher than a linear fit. For an ultracold molasses ($T \sim 100~\mu$K) we see no statistically significant difference between the quadratic and linear fits at the 4\% level. For the data in Figs.~\ref{MOT} and \ref{TvsI} we use the quadratic model, while for the data in Fig.~\ref{Molasses} we use the linear model.

The magnification of the imaging system may not be perfectly uniform across the field of view. This can alter the apparent size of the cloud as it drops under gravity and expands. We have measured the magnification across the whole field of view that is relevant to our data and find that the uniformity is better than 3\%. At this level, the effect on the temperatures is negligible.

Finally, a non-uniform magnetic field can result in an expansion that does not accurately reflect the temperature. A magnetic field gradient accelerates the molecules, and since the magnetic moment depends on the hyperfine component and Zeeman sub-level this acceleration is different for different molecules. We calculate that this effect contributes a velocity spread less than 0.5~cm/s after 10~ms of free expansion. This is negligible, even for a 50~$\mu$K cloud. The second derivative of the magnetic field causes a differential acceleration across the cloud, but this effect is even smaller.

\end{document}